\documentclass[10pt,twocolumn,letterpaper]{article}
\usepackage{fancyhdr} 
\usepackage{iccv}
\usepackage{times}
\usepackage{graphicx}
\usepackage{amsmath}
\usepackage{amssymb}
\usepackage{booktabs}

\usepackage[pagebackref=true,breaklinks=true,letterpaper=true,colorlinks,bookmarks=false]{hyperref}

\iccvfinalcopy 

\def\iccvPaperID{****} 

\ificcvfinal\fi

\begin{document}

\title{IDOL-Net: An Interactive Dual-Domain Parallel Network for CT Metal Artifact
Reduction}

\author{Tao Wang\(^1\), Wenjun Xia\(^1\), Zexin Lu\(^1\), Huaiqiang Sun\(^2\), Yan Liu\(^3\), Jiliu Zhou\(^1\), Yi Zhang\(^{1*}\)\\
\(^1\)College of Computer Science, Sichuan University, Chengdu, China\\
\(^2\)Department of Radiology, West China Hospital of Sichuan University, Chengdu, China\\
\(^3\)College of Electrical Engineering, Sichuan University, Chengdu, China\\
{\tt\small \(scuer\_wt\)\(@stu.scu.edu.cn\) \ \(xwj90620@gmail.com\) \ \(zexinlu.scu@gmail.com \) }\\
{\tt\small\(\{sunhuaiqiang, liuyan77, huchen, zhoujl, yzhang\}@scu.edu.cn\)}

}

\maketitle
      \thispagestyle{fancy} 
      \lhead{} 
      \chead{} 
      \rhead{} 
      \lfoot{} 
      \cfoot{} 
      \rfoot{\thepage} 
      \renewcommand{\headrulewidth}{0pt} 
      \renewcommand{\footrulewidth}{0pt} 
\pagestyle{fancy}
\rfoot{\thepage}
\ificcvfinal\thispagestyle{empty}\fi

\begin{abstract}
   Due to the presence of metallic implants, the imaging quality of computed tomography (CT) would be heavily degraded. With the rapid development of deep learning, several network models have been proposed for metal artifact reduction (MAR). Since the dual-domain MAR methods can leverage the hybrid information from both sinogram and image domains, they have significantly improved the performance compared to single-domain methods. However, current dual-domain methods usually operate on both domains in a specific order, which implicitly imposes a certain priority prior into MAR and may ignore the latent information interaction between both domains. To address this problem, in this paper, we propose a novel interactive dual-domain parallel network for CT MAR, dubbed as IDOL-Net. Different from existing dual-domain methods, the proposed IDOL-Net is composed of two modules. The disentanglement module is utilized to generate high-quality prior sinogram and image as the complementary inputs. The follow-up refinement module consists of two parallel and interactive branches that simultaneously operate on image and sinogram domain, fully exploiting the latent information interaction between both domains. The simulated and clinical results demonstrate that the proposed IDOL-Net outperforms several state-of-the-art models in both qualitative and quantitative aspects.
\end{abstract}

\section{Introduction}
Computed tomography (CT) has been widely applied in clinical, industrial and other areas ~\cite{ref1}. However, when metallic implants exist in scanning regions, the effects of beam hardening, photon starvation, scattering and nonlinear partial volume effect become prominent, giving rise to severe radial streak artifacts in reconstructed CT images, which may seriously affect the subsequent analysis or diagnosis tasks ~\cite{ref2}. As a result, it is of great importance to develop an efficient method to reduce metal artifacts while preserving tissue details. 
\begin{figure}
\begin{center}
\centering\includegraphics[width=0.45\textwidth]{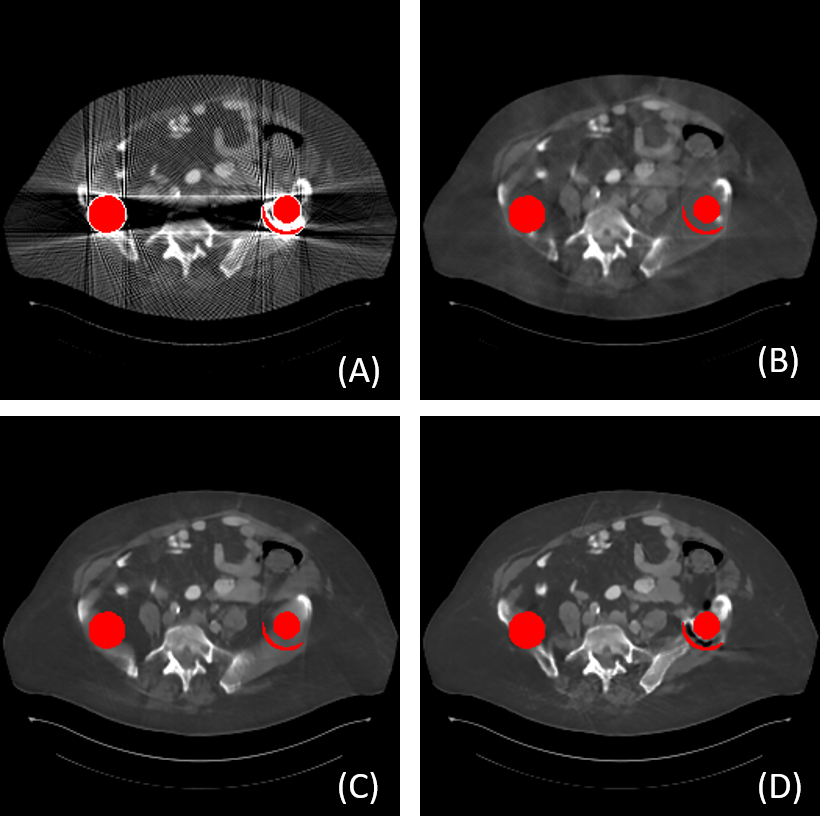}
\end{center}
   \caption{A representative slice processed using different MAR methods. (A) Metal-corrupted image, (B) CNNMAR, (C) DuDoNet, (D) IDOL-Net}
   \label{fig0}
\end{figure}
\section{Method}
\subsection{Overview}
During the past decades, many metal artifact reduction (MAR) methods have been proposed ~\cite{ref3}. These methods can be roughly divided into three groups: sinogram completion, iterative reconstruction and image post-processing. 1) Sinogram completion methods regard projection data in the metal trace region as missing, which are estimated using image interpolation or inpainting methods ~\cite{ref4}~\cite{ref5}~\cite{ref6}~\cite{ref7}~\cite{ref8}~\cite{ref9}~\cite{ref10}~\cite{ref11}. This kind of method is difficult to maintain continuity at the boundaries of metal trace in sinogram domain and secondary artifacts will be introduced in the reconstructed CT images ~\cite{ref12}. 2) Iterative reconstruction methods usually formulate the objective function aided by prior information, which can be solved with an iterative manner ~\cite{ref13}~\cite{ref14}~\cite{ref15}~\cite{ref16}. Nevertheless, these approaches are usually time-consuming [3] and some parameters need to be set manually. 3) Image post-processing methods directly remove artifacts in reconstructed image ~\cite{ref17}~\cite{ref18}, but they have to entail risks of distorting the anatomic structure ~\cite{ref19}.

Recently, increasing attention has been paid to deep learning (DL). As a result, many DL-based MAR approaches were proposed, which also can be grouped into three different categories according to their operation domain. For sinogram-domain methods, projection data in metal trace region is replaced with the values generated by neural networks. Image-domain methods dedicate to suppressing metal artifacts directly without accessing projection data. In the last few years, dual-domain methods achieve impressive performance, which simultaneously leverage the information from both image and sinogram domains. However, current dual-domain methods usually process dual-domain data in a specific order, which implicitly imposes a certain priority prior into MAR and may ignore the latent information interaction between both domains. On the other side, as shown in recent studies \cite{ref19}~\cite{ref20}~\cite{ref21}~\cite{ref32}, it is preferred to produce a prior image as the auxiliary input to improve the performance of networks. As a result, how to obtain a high-quality initial estimation becomes another critical point for DL-based MAR methods.

To tackle these problems mentioned above, in this paper, we propose a novel Interactive Dual-dOmain paraLlel network for CT MAR, dubbed as IDOL-Net. The proposed IDOL-Net is composed of two modules as depicted in Fig. 1. The disentanglement module introduces the idea of disentanglement representation to generate high-quality prior sinogram and image as the complementary inputs. The follow-up refinement module consists of two parallel and interactive branches that simultaneously operate on image and sinogram domain, attempting to fully exploit the latent information interaction between both domains. The intermediate results exported from both branches are fused to exchange the useful features from each other. Finally, the image quality is further enhanced by the refinement module. A representative slice processed using different MAR methods is given in Fig. \ref{fig0}.

The main contributions of this work are summarized as follows:

(1)	A novel dual-domain network, IDOL-Net, which can simultaneously leverage the information from both sinogram and image domains, is proposed to estimate a complex end-to-end mapping from joint metal-affected sinogram and image to artifact-free image.

(2)	Disentanglement representation is introduced to generate a high-quality hybrid prior estimation from the initial metal-affected sinogram and reconstructed image.

(3)	A refinement module containing two parallel and interactive branches is designed to simultaneously extract the intermediate features from both complementary domains, which are delivered and fused during the reconstruction procedure.

(4)	Extensively experimental results on the simulated and clinical datasets demonstrate the effectiveness compared to other state-of-the-art MAR methods in both qualitative and quantitative aspects.

\section{Related works}
\subsection{Single-domain DL-based MAR}
Early MAR methods directly applied DL on single-domain data, i.e., sinogram or image reconstructed by filter backprojection (FBP). For sinogram domain, neural networks are utilized to synthesize the missing data in the metal trace region. Park et al. used U-Net ~\cite{ref23} to repair inconsistent sinograms by removing the primary metal-induced beam hardening factors along the metal trace boundaries ~\cite{ref22}. In ~\cite{ref24}~\cite{ref25}~\cite{ref26}, generative adversarial net (GAN) and its variant were introduced to improve the inpainting performance. Partial convolution ~\cite{ref27}, which is more suitable to sinogram completion, was employed in ~\cite{ref28} and ~\cite{ref29}. However, it is hard to guarantee the sinogram continuity at the boundaries of metal trace and secondary artifacts will appear consequentially. On the other hand, Gjesteby et al. first applied a simple plain CNN to suppress the artifacts in image domain ~\cite{ref30}. Zhang et al. proposed to fuse the results obtained using different methods as the prior image and replace the metal-affected region in the sinogram with the corresponding part from the prior image ~\cite{ref34}. In ~\cite{ref31}, the authors presented to combine the result of NMAR ~\cite{ref8} and the detail image extracted by guided filter as hybrid inputs to a dual-stream network. Liao et al. introduced the idea of unsupervised disentanglement representation to recover the clean image with unpaired training samples ~\cite{ref36}. The main limitation for this method lies in that when the metallic implants get larger, the artifacts cannot be completely removed and the anatomic structures are prone to distortion.

\subsection{Dual-domain DL-based MAR}
Recently, dual-domain methods become mainstream. For instance, in ~\cite{ref20}, two sub-networks are jointed via a Radon inversion layer to process the linear-interpolated sinogram and its reconstructed image in sequence. To further improve the performance, the authors in ~\cite{ref21} proposed to insert metal mask projection into the sinogram network. Yu et al. pointed out that there would be some tiny anatomical structure changes if image domain enhancement is adopted at the final stage ~\cite{ref19}. Thus, an image domain network is trained at first to produce a prior image with good quality to guide the subsequent sinogram domain learning. Peng et al. utilized partial convolution operator to recover the contaminated values in sinogram and trained an auxiliary sub-network in image domain to further refine the details ~\cite{ref32}. However, since these methods usually process the sinogram and image in a specific order, which ignores the latent information interaction, there is still room to improve the performance by searching for a better network architecture for dual-domain information fusion.
\begin{figure*}
\begin{center}
\centering\includegraphics[width=1\textwidth]{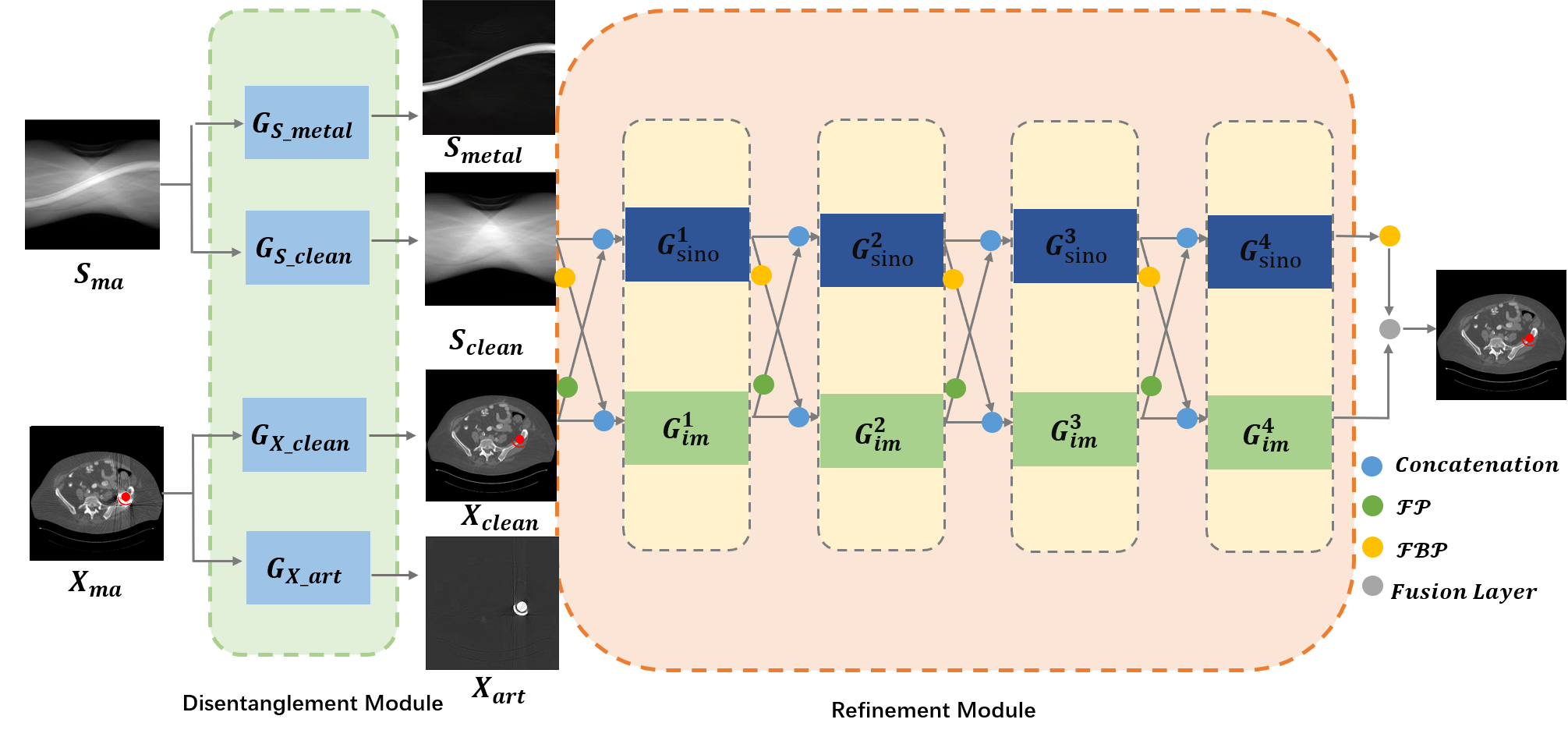}
\end{center}
   \caption{Overview of the proposed MAR method.}
   \label{fig1}
\end{figure*}
\section{Method}
\subsection{Overview}
As depicted in Fig. \ref{fig1}, the proposed IDOL-Net is composed of two modules, the disentanglement module and refinement module. In the disentanglement module, we try to obtain accurate prior estimations, including both initial sinogram and image. In the refinement module, the preliminarily corrected sinogram and image from disentanglement module are further refined aided by interactive feature fusion. The details of both modules are elaborated as follows.

\subsection{Disentanglement module}
Current DL-based methods ~\cite{ref19}~\cite{ref20}~\cite{ref21}~\cite{ref34} have demonstrated the effectiveness in introducing the prior image or sinogram separately as the auxiliary input for MAR. However, although it is natural to combine both image and sinogram as the prior inputs, none of these methods have explored this possibility. To take full advantage of the prior information from both domains, unlike other DL-based methods, which use linear interpolation or U-Net to obtain the coarse prior sinogram or image, the proposed disentanglement module introduces the idea of disentanglement representation to produce a more accurate hybrid prior estimation.

\begin{figure*}
\begin{center}
\centering\includegraphics[width=1\textwidth]{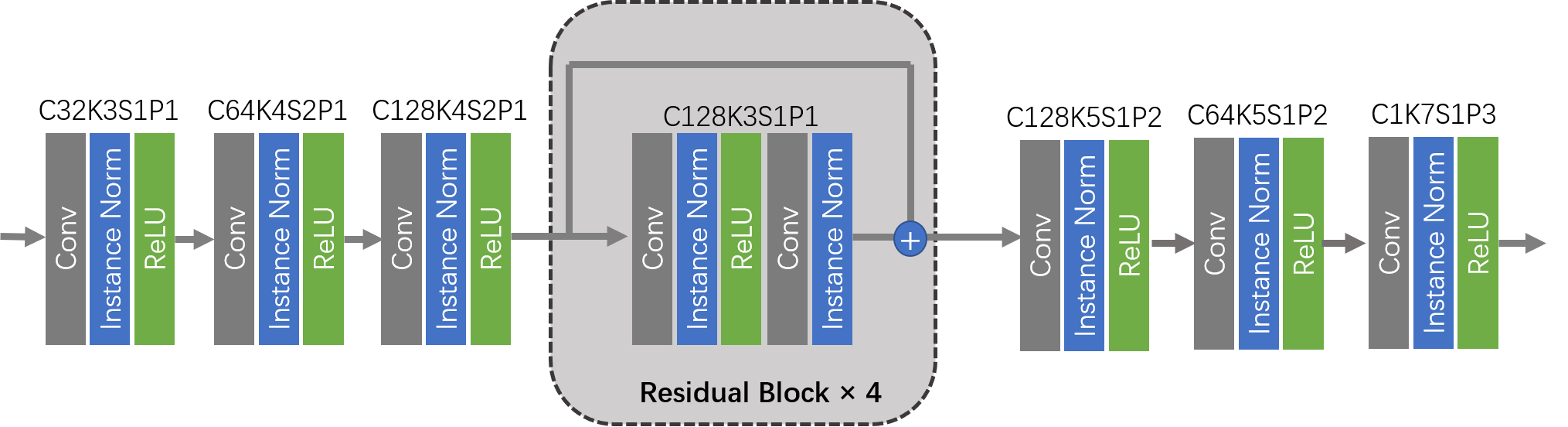}
\end{center}
\caption{Fig. 2: The details of \(G_{S\_metal}\) , \(G_{S\_clean}\)  , \(G_{X\_art}\) and \(G_{X\_clean}\) . C: channel numbers, K: kernel, S: stride, and P: padding sizes.}
   \label{fig2}
\end{figure*}

As illustrated in the left part of Fig. \ref{fig1}, the original metal-affected sinogram \(S_{ma}\) is fed into two networks, \(G_{S\_metal}\) and \(G_{S\_clean}\), which respectively extract the metal-only sinogram \(S_{metal}\) and the content sinogram \(S_{clean}\)  from \(S_{ma}\):
\begin{equation}
\label{eq1}
S_{metal}= G_{S\_metal} \ (S_{ma} )
\end{equation}\begin{equation}
\label{eq2}
S_{clean}= G_{S\_clean}\ (S_{ma} )
\end{equation}
Meanwhile, the artifact-corrupted image \(X_{ma}\) reconstructed from \(S_{ma}\) using FBP is also processed by two networks, \(G_{X\_art}\)  and \(G_{X\_clean}\) , which respectively estimate the artifact image \(X_{art}\)  and the artifact-free image \(X_{clean}\) :
\begin{equation}
\label{eq3}
X_{art}= G_{X\_art}\  (X_{ma} )
\end{equation}
\begin{equation}
\label{eq4}
X_{clean} = G_{X\_clean}\   (X_{ma} )
\end{equation}Unlike the traditional MAR networks, which usually utilize one network to generate clean sinogram or image, the proposed disentanglement module imposes additional constraints on the extracted residuals, i.e., the metal-only sinogram and the artifact image, producing a purer decomposition on the original sinogram and image. The loss function of the disentanglement module is defined as:
\begin{equation}
\label{eq5}
\mathcal L_{dis}=\mathcal L_{clean}+\alpha_{1}\mathcal L_{art}+\alpha_2\mathcal L_{ma}
\end{equation}
Three components are formulated respectively as
\begin{equation}
\label{eq6}
\mathcal L_{clean}=||X_{clean}-X_{clean}^{gt}||_{1}+\theta_{1}||S_{clean}-S_{clean}^{gt}||_{1}
\end{equation}
\begin{equation}
\label{eq7}
\mathcal L_{art}=||X_{art}-X_{art}^{gt}||_{1}+\theta_{2}||S_{metal}-S_{metal}^{gt}||_{1}
\end{equation}
\begin{equation}
\begin{aligned}
\label{eq8}
\mathcal L_{ma}&=||(X_{clean}+X_{art})-X_{ma}||_{1}\\&+\theta_{3}||(S_{clean}+S_{metal})-S_{ma}||_{1}
\end{aligned}
\end{equation}
where \(X_{clean}^{gt}\) , \(S_{clean}^{gt}\) , \(X_{art}^{gt} \) and \(S_{metal}^{gt}\)  are the corresponding labels. \(\mathcal{L}_{clean}\) denotes the reconstruction loss to facilitate \(X_{clean}\approx X_{clean}^{gt}\) and \(S_{clean}\approx S_{clean}^{gt}\); \(\mathcal{L}_{art}\) indicates the reconstruction loss to impose \(X_{art}\approx X_{art}^{gt}\) and \(S_{metal}\approx S_{metal}^{gt}\); \(\mathcal{L}_{ma}\) is adopted to ensure that the decomposition is invertible and without loss of information.
For simplicity, in this paper, \(G_{S\_metal}\) , \(G_{S\_clean}\)  , \(G_{X\_art}\)  and \(G_{X\_clean}\)  share the same structure, which is presented in Fig. \ref{fig2}. A simple plain CNN with four residual blocks is adopted as the unified network architecture.

\subsection{Refinement module}
\begin{figure}[t]
\begin{center}
 \includegraphics[width=1\linewidth]{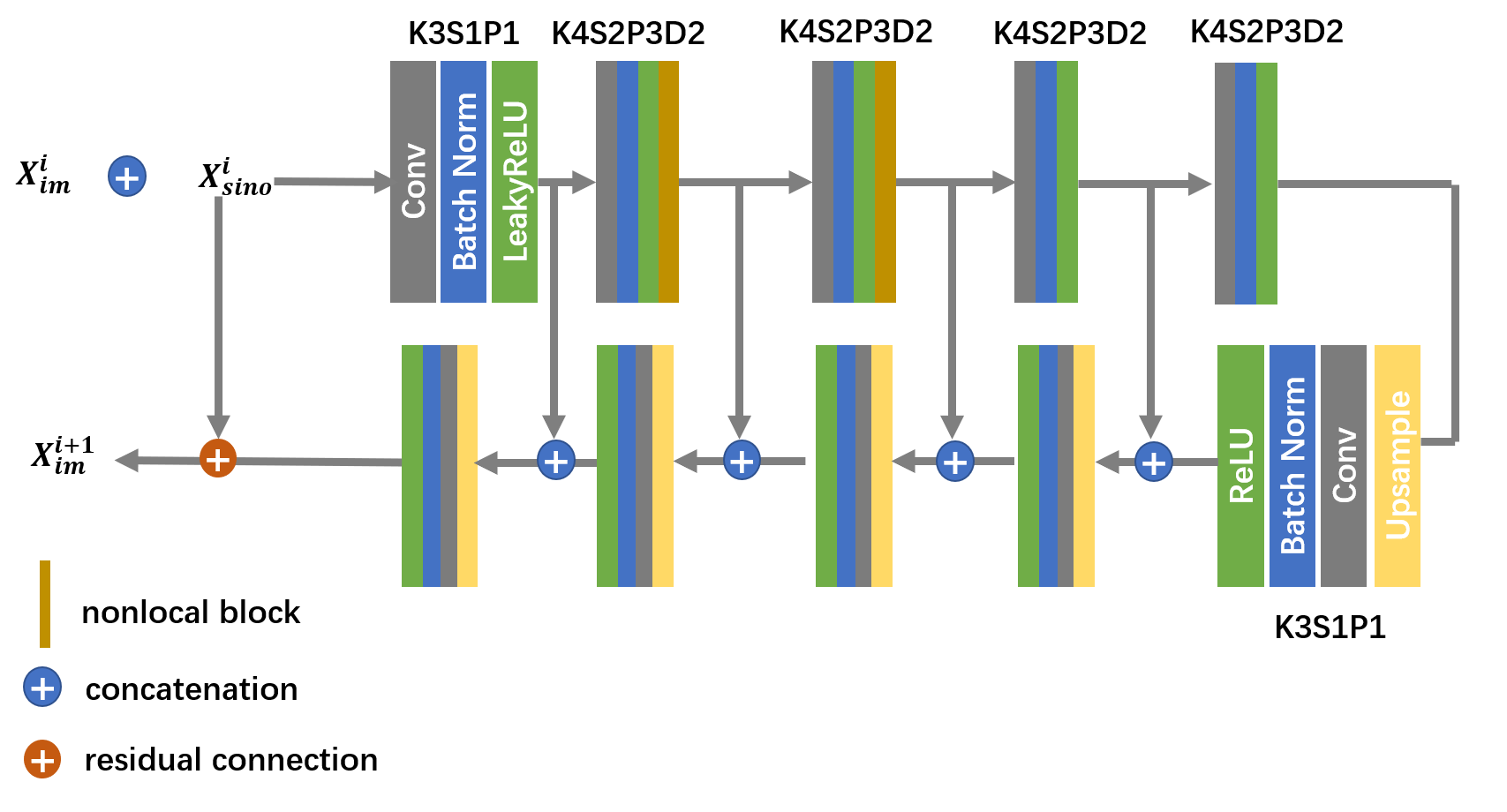}
\end{center}
   \caption{Illustrations of the \(G_{im}\). K: kernel, S: stride, P: padding sizes and D: dilation.}
      \label{fig3}
\end{figure}
\begin{figure}[t]
\begin{center}
 \includegraphics[width=1\linewidth]{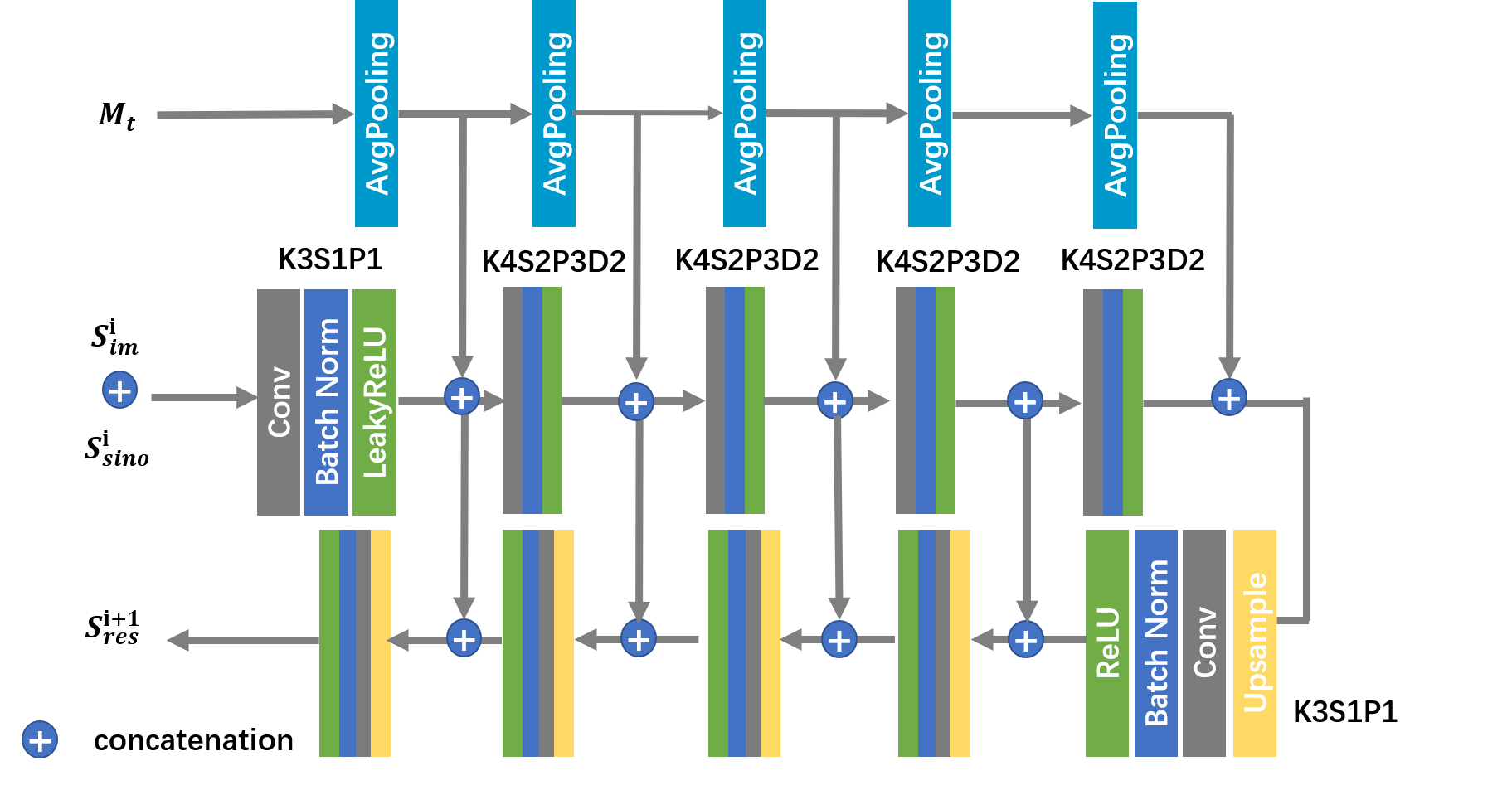}
\end{center}
   \caption{Illustrations of the \(F_{sino}\). K: kernel, S: stride, P: padding sizes and D: dilation.}
      \label{fig4}
\end{figure}
Existing studies demonstrate that simply processing the corrupted sinogram or image with one single network has limited performance ~\cite{ref19}~\cite{ref20}. Based on this consideration, a refinement module is proposed to further improve the image quality. Its structure is shown in the right part of Fig. \ref{fig1}. There are two branches to separately handle sinogram and image data. Both branches have four same blocks, i.e., \(G_{im}^{i}\) for image branch and \(G_{sino}^{i}\) for sinogram branch. \(X_{clean}\) and \(S_{clean}\) generated by the disentanglement module are employed as the inputs of the proposed refinement module. FBP is applied on \( S_{clean}\) to obtain \(X_{clean}^{fbp}\), which is concatenated with \(X_{clean}\) as the initial input of the first \(G_{im}^{1}\). Similarly, forward projection (FP) is applied on \(X_{clean}\) to obtain \(S_{clean}^{fp}\), which is concatenated with \(S_{clean}\) as the initial input of the first \(G_{sino}^{i+1}\) . Similar operations are performed after each \(G_{im}^{i}\) and \(G_{sino}^{i}\). FBP and FP are applied to the intermediate sinogram and image results. Then the outputs of both FBP and its corresponding \(G_{im}^{i}\) are concatenated and fed into \(G_{im}^{i+1}\). The outputs of both FP and its corresponding \(G_{sino}^{i}\) are concatenated and fed into  \( G_{sino}^{i+1}\). For simplicity, U-Net is adopted as the backbone of both \(G_{sino}^{i}\) and \(G_{im}^{i}\) and we halve the channel numbers to reduce the computational cost. To suppress the nonlocal artifacts in image domain, nonlocal network modules ~\cite{ref37} are integrated into \(G_{im}^{i}\). The architecture of \(G_{im}^{i}\) is illustrated in Fig. \ref{fig3}. For sinogram branch, a sinogram completion network, denoted as \(F_{sino}^{i}\), is adopted to restore projection data within the metal trace mask and the structure of \(F_{sino}^{i}\) is depicted in Fig. \ref{fig4}. Inspired by the idea of mask pyramid network (MPN) ~\cite{ref33}, the metal mask projection is leveraged to alleviate discontinuity at boundaries of metal trace.

Specifically, in a refinement block, given the metal mask \(M \) and metal trace mask \(M_t\), there are four outputs, outputs of \(G_{sino}^{i+1}\) and \(G_{im}^{i+1}\) recognized as \(S_{sino}^{i+1}\)  and \(X_{im}^{i+1}\), forward projection of \(X_{im}^{i+1}\) denoted as \(S_{im}^{i+1}\) and filtered back-projection of \(S_{sino}^{i+1}\) referred to as \(X_{sino}^{i+1}\) . 
\begin{equation}
\label{eq9}
S_{res}^{i+1} = F_{sino}^{i+1}(S_{im}^{i}\ ,S_{sino}^{i}\ ,M_{t})
\end{equation}
\begin{equation}
\begin{aligned}
\label{eq10}
S_{sino}^{i+1} &= G_{sino}^{i+1}(S_{res}^{i+1} ,S_{ma} ,M_{t})\\&=S_{res}^{i+1}\odot M_{t}+S_{ma}\odot(1-M_{t})
\end{aligned}
\end{equation}
\begin{equation}
\label{eq11}
X_{im}^{i+1} = G_{im}^{i+1}(X_{im}^{i}\ ,X_{sino}^{i})+X_{im}^{i}
\end{equation}
\begin{equation}
\label{eq12}
S_{im}^{i+1} = \mathcal{FP} (X_{im}^{i+1})
\end{equation}
\begin{equation}
\label{eq13}
X_{sino}^{i+1} = \mathcal{FBP} (S_{sino}^{i+1})
\end{equation}
where i=0,1,2,3. When i=0, \(S_{sino}^{1}=S_{clean}\), \(S_{im}^{1}=S_{clean}^{fp}\), \(X_{im}^{1}=X_{clean}\) and \(X_{sino}^{1}=X_{clean}^{fbp}\).
The first three refinement blocks share the same structure. The last refine block has similar architecture, but it has no FP operation on \(X_{im}^{4}\). Then, we concatenate \(X_{im}^{4}\) and \(X_{sino}^{4}\)  of the last block as the inputs of the fusion layer \(\mathcal{F}\), which is a \(1\times1\) convolution layer, to obtain the final result \(X_{corr}\) as 
\begin{equation}
\label{eq14}
X_{corr} = \mathcal{F} (X_{im}^{4}\ ,X_{sino}^{4})
\end{equation}

To optimize the refinement module, \(\mathcal{L}_1 \) loss is adopted to minimize the differences between  \(X_{corr} \) and the ground truth:
\begin{equation}
\label{eq15}
\mathcal{L}_{corr} = ||(X_{corr}-X_{gt}^{clean})\odot (1-M_{t})||_1
\end{equation}Thus, the total loss function can be written as:

\begin{equation}
\label{eq16}
\mathcal{L} = \mathcal{L}_{dis}+\gamma\mathcal{L}_{corr}
\end{equation}
where \(\alpha_1\), \(\alpha_2\), \(\theta_1\),\( \theta_2\), \(\theta_3\) and \(\gamma\) are hyper-parameters to balance the weights of different components. We manually set \(\alpha_1=\alpha_2=1.0\) , \(\theta_1=\theta_2=\theta_3=0.06\) and \(\gamma=10.0\) in this manuscript.
\section{Experiments}
\subsection{Data set}
In our experiments, the manually segmented metal masks dataset ~\cite{ref34}, which contains simulated dental fillings, spine crews and hip prostheses, is utilized to produce the metal masks. We randomly selected 1000 CT images from DeepLesion dataset ~\cite{ref35} and 90 metal masks to build our training set, which contains totally 90000 images. Another 200 CT images and ten masks were used to form the testing set with totally 2000 images. The same procedure used in ~\cite{ref34}~\cite{ref36} is adopted to synthesize the metal-affected CT images. Images were resized to 256×256 to reduce the computational cost. Without loss of generality, 2D parallel-beam geometry was employed. There were 361 views evenly distributed from 0° to 180° and 367 bins received photons. In addition, the pixel values were clipped to [0, 4095], which coincided with the real scenario.
\subsection{Implementation details}
The \textit{ODL} library was adopted to implement FBP and FP. The whole network was trained in an end-to-end manner with the \textit{Pytorch} deep learning framework and was optimized using the Adam optimizer with the parameters \((\beta_{1},\beta_{2})=(0.5,0.999)\). The learning rate was set to 0.0002 initially and halved for every 10 epochs. We trained our network for 100 epochs on a workstation with four Nvidia 1080Ti GPUs.

\begin{table*}[]
\centering
\label{table1}
\begin{tabular}{@{}cccccccc@{}}
\toprule
\hline
PSNR/SSIM & 1&2     &3   &4  &5 &Average\\ 
\midrule
\hline
Uncorrected & 15.93/0.7101 & 15.14/0.6617 & 14.86/0.6417 & 15.14/0.6572 & 15.32/0.6564 & 15.33/0.6673 \\
LI        & 32.86/0.9545 & 32.60/0.9442 & 30.76/0.9330 & 27.26/0.8934 & 28.94/0.8853 & 30.74/0.9224 \\
NMAR      & 32.54/0.9564 & 31.87/0.9394 & 31.20/0.9359 & 27.91/0.9035 & 29.21/0.9000 & 30.83/0.9270 \\
CNNMAR    & 32.78/0.9665 & 32.80/0.9641 & 33.24/0.9612 & 32.61/0.9517 & 32.46/0.9455 & 32.72/0.9574 \\
ADN       & 30.92/0.9335 & 30.03/0.9162 & 32.08/0.9393 & 26.45/0.9136 & 31.78/0.9407 & 30.58/0.9290 \\
DuDoNet   & 38.77/0.9673 & 38.57/0.9648 & 36.44/0.9642 & 33.03/95.33 &34.26/0.9531 & 36.82/0.9777 \\
IDOL-Net  & \textbf{42.01/0.9870} &\textbf{ 42.37/0.9878}  & \textbf{40.67/0.9900} & \textbf{35.26/0.9782 }& \textbf{38.30/0.9842} & \textbf{40.20/0.9858}\\
\bottomrule
\hline
\end{tabular}
\caption{Quantitative comparison of different methods on the simulated dataset. 1-5 represents the sizes of metallic implants in ascending order.}
\end{table*}

\begin{figure*}
\begin{center}
\centering\includegraphics[width=0.8\textwidth]{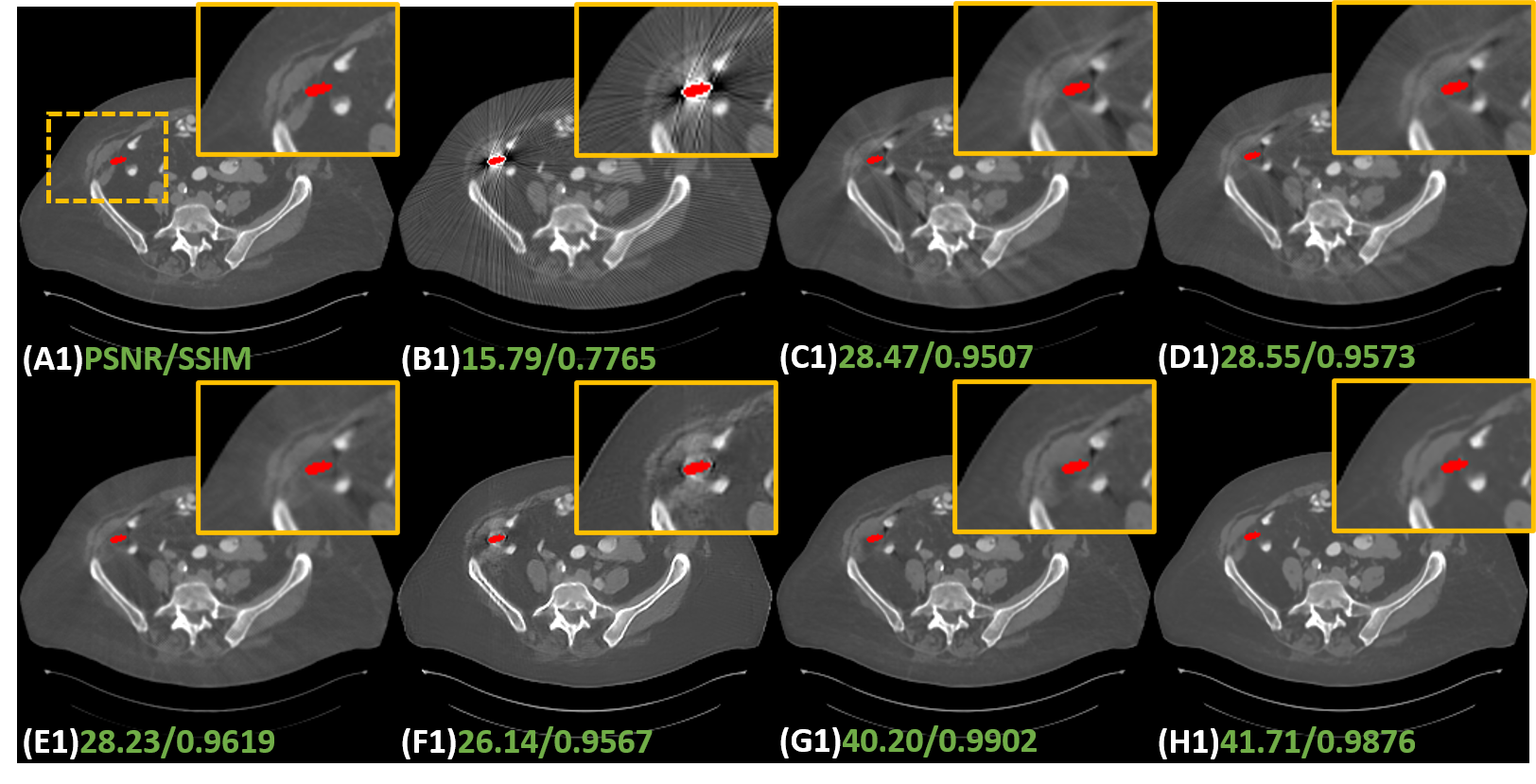}
\centering\includegraphics[width=0.8\textwidth]{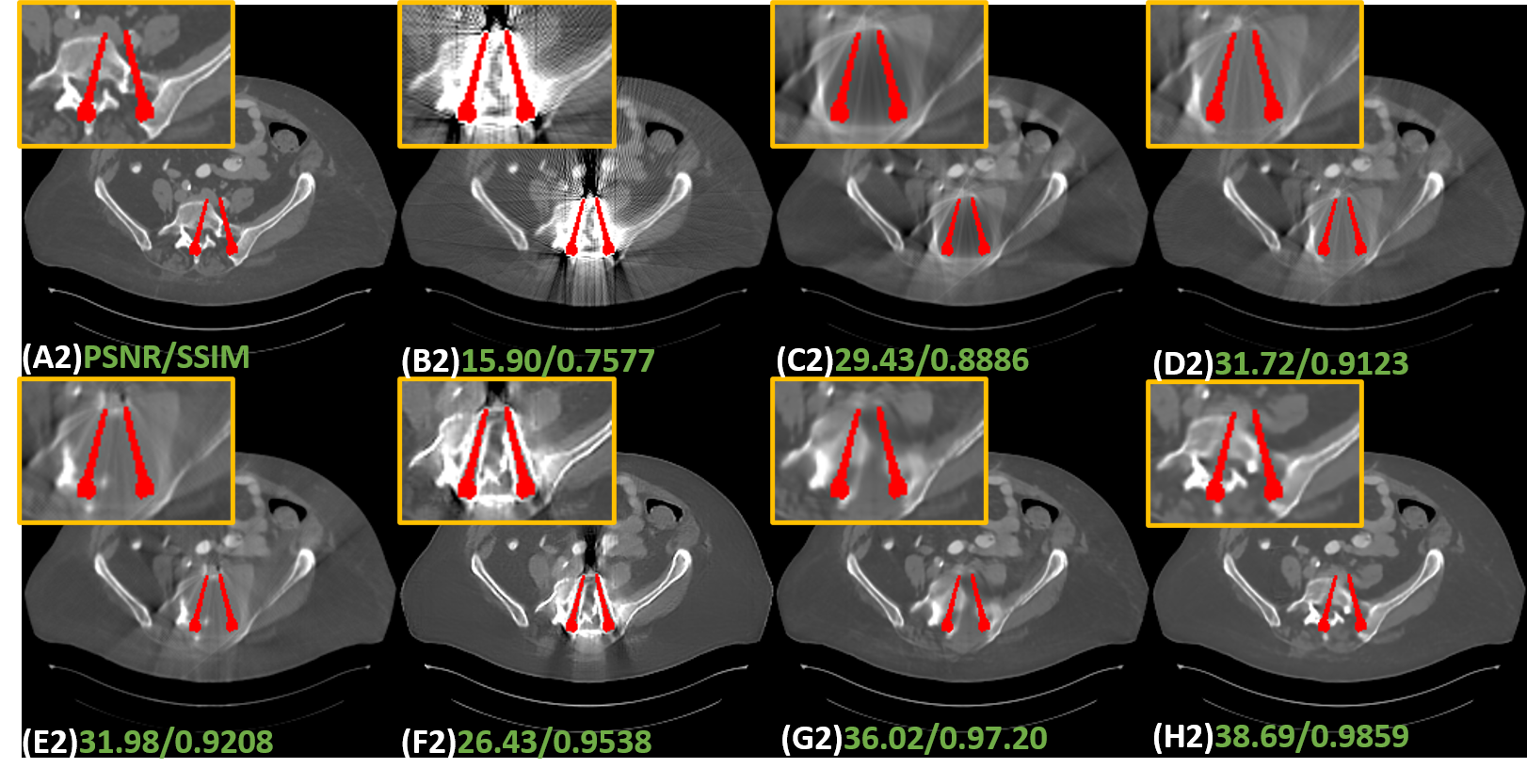}
\centering\includegraphics[width=0.8\textwidth]{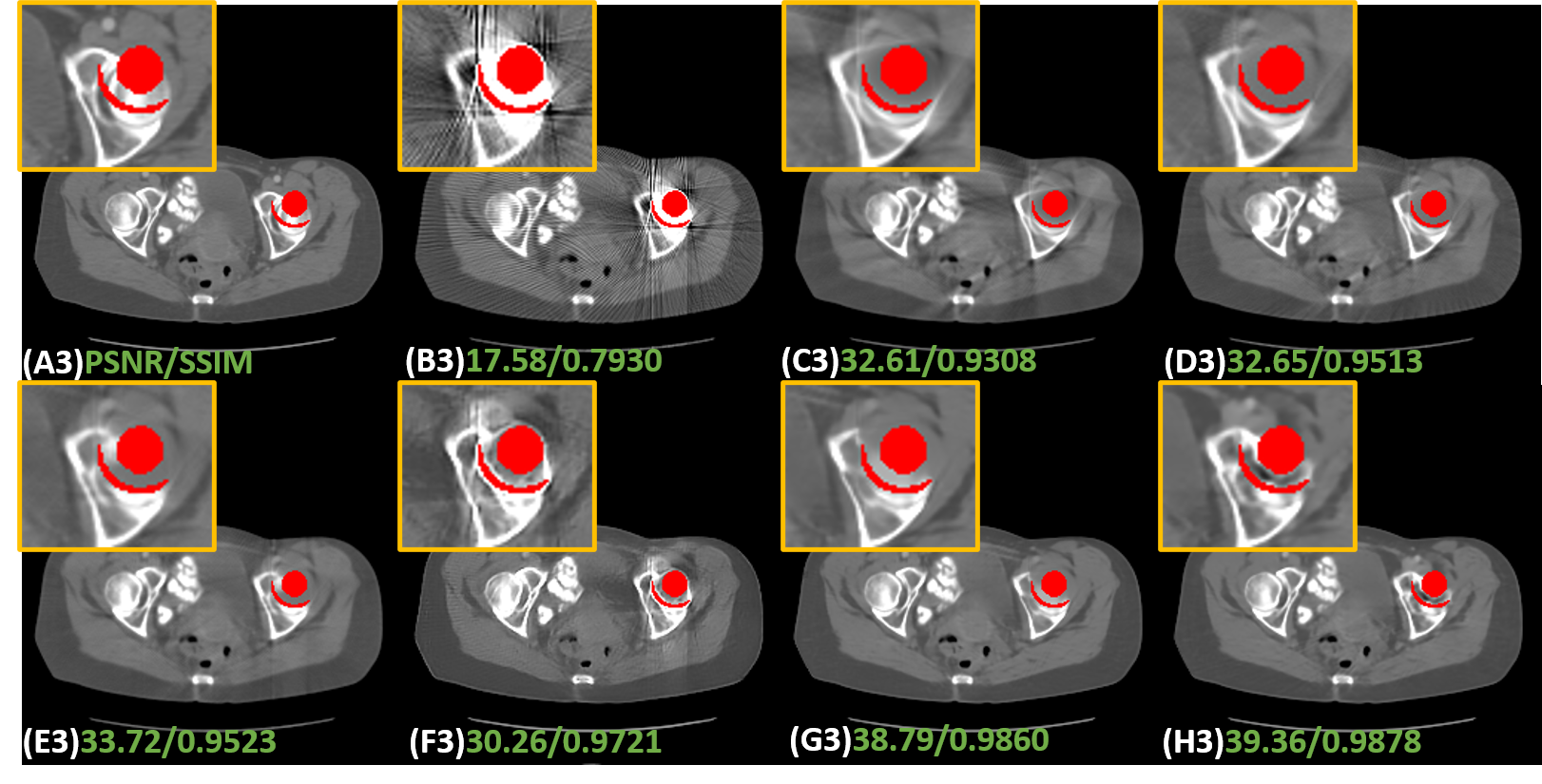}
\end{center}
   \caption{Comparisons with different MAR methods on simulation data with small, middle and big metallic implant. (A1-A3) reference images; (B1-B3) metal corrupted images; (C1-C3) corresponding results of LI; (D1-D3) NMAR; (E1-E3) CNNMAR; (F1-F3) ADN; (G1-G3) DuDoNet and (H1-H3) IDOL-Net.}
   \label{fig5}
\label{fig:short}
\end{figure*}

\begin{table*}[]
\centering
\label{table2}
\begin{tabular}{@{}cccccccc@{}}
\toprule
\hline
Methods & Refine-Net & Clean-IDOL  & Art-IDOL &Prior-Im   &Prior-Sino &IDO-Net\\ 
\midrule
\hline
PSNR & 36.35 & 36.65 &37.52  &31.42  &27.30 & \textbf{40.20} \\
SSIM &0. 9660 & 0.9716 & 0.9723 &0.9418   &0.6005  & \textbf{0.9858} \\
\bottomrule
\hline
\end{tabular}
\caption{Quantitative comparison of different variants of our method on the simulated dataset.}
\end{table*}

\subsection{Experimental results on the simulated data set}

To validate the effectiveness of the proposed IDOL-Net, several methods, including LI ~\cite{ref6}, NMAR ~\cite{ref8}, CNNMAR ~\cite{ref34}, ADN ~\cite{ref36}, DuDoNet ~\cite{ref20} were involved for comparison. LI and NMAR are widely used interpolation-based methods. CNNMAR takes advantage of the outputs of different MAR methods to generate a prior image. ADN is an advanced unsupervised image-domain method. DuDoNet is a state-of-the-art dual-domain approach. All the methods were implemented with the publicly released code or strictly implemented according to the original paper. Structural similarity (SSIM) and peak signal to noise ratio (PSNR) were employed as the quantitative metrics.

\textbf{1) Quantitative comparisons:}  Table \ref{table1} shows the average quantitative results on the simulated dataset. It is observed that all the methods can significantly improve the PSNR and SSIM values and DL-based methods achieve better scores than traditional MAR methods. DuDoNet and IDOL-Net gain obvious improvements in terms of both metrics, which quantitatively demonstrated the merits of dual-domain methods. In addition, our method produces a noticeable advantage over DuDoNet in terms of both metrics for all the metal sizes, with an average improvement of 3.3dB in PSNR and 0.81 in SSIM.

\textbf{2) Qualitative comparisons:} Fig. \ref{fig5} shows the representative results of different methods on simulated data with different metal sizes. For better visualization, the metal masks are painted in red. Due to the fact that LI and NMAR are interpolation-based methods directly discarding the projection data in metal trace, the metal information is lost and the corrected sinogram cannot keep the continuity at the boundaries of metal trace, leading to blurred tissues around the metals and remarkable secondary artifacts. Although CNNMAR fuses the results of different MAR methods, the improvement is quite limited. ADN achieves better performance on structure recovery around the metal implants than LI, NMAR and CNNMAR, and even DuDoNet in some cases, but it also can be noticed that in the second case (Fig. \ref{fig5} (F2)), the shadow artifacts are not well suppressed and the details covered by the artifacts are hard to identify. This phenomenon probably lies in that ADN is essentially a post-processing method, which does not leverage the sinogram information. Compared to other methods, DuDoNet and the proposed IDOL-Net have the best performance on artifact reduction, but DuDoNet cannot recover the structures and tissues around the metal well, which can be clearly observed in the magnified yellow boxes in Fig. \ref{fig5} (G2) and (G3). Our proposed IDOL-Net not only effectively removes most artifacts but also recovers the details better than all the other methods in all cases.
\subsection{Experimental results on clinical data}
\begin{figure}[]
\begin{center}
\centering\includegraphics[width=0.5\textwidth]{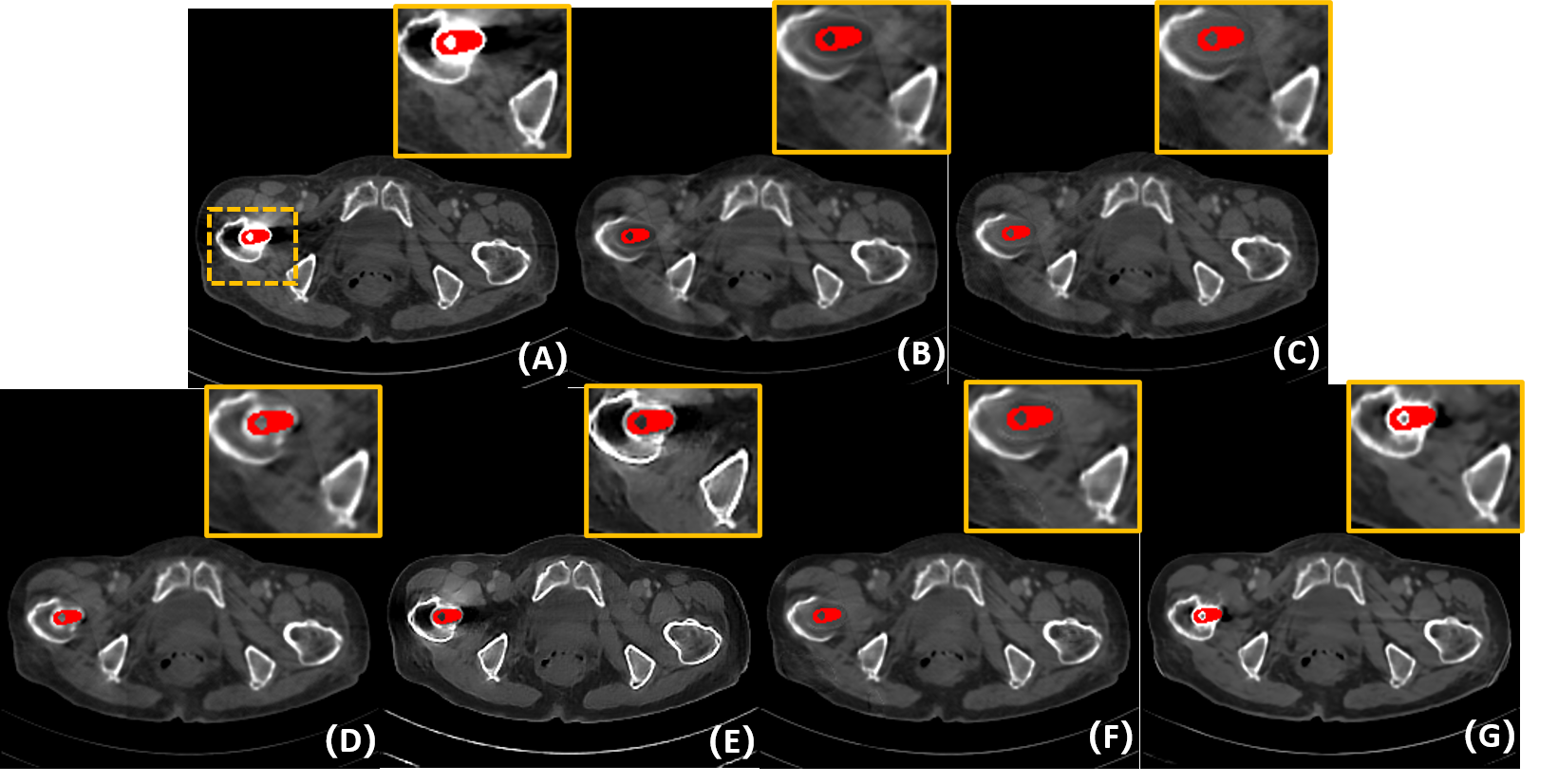}
\end{center}
   \caption{Visual comparison with different MAR methods on a clinical CT image. (A)-(G) respectively represent uncorrected CT images and corrected results using LI, NMAR, CNNMAR, ADN, DuDoNet and IDOL-Net.}
   \label{fig6}
\end{figure}

To evaluate the robustness of our IDOL-Net for clinical practice, clinical CT images were tested and Fig. \ref{fig6} shows the visual comparisons of one representative slice processed using different methods. The metal was segmented by the threshold of 2000 HU and the metal mask is painted in red for better visualization. The test image was normalized to the same range as the training data and the results were obtained using the model trained with simulated dataset used in the previous subsection. In Fig. \ref{fig6} (A), the metal leads to severe artifacts, which covers some tissues, especially near the metal. It is observed that LI, NMAR and DuDoNet cannot retrieve the tissues around the metal, but due to the utilization of dual-domain information, DuDoNet outperforms most methods in artifact reduction. Since the shadow artifacts around the metal is very heavy, ADN cannot do much about this, which is similar to the situation in simulated dataset. It is observed that IDOL-Net suppresses most of the metal artifacts and effectively preserves the anatomical structures around the metals, which demonstrates its potential for real clinical application.


\subsection{Analytical studies}
In this subsection, we investigate the effectiveness of different network modules and our ablation experiments use the following configurations.
(a): Refine-Net: network trained without both prior image and sinogram;

(b): Clean-IDOL: network that only learns clean parts of CT images and sinograms followed by refinement module;

(c): Art-IDOL: network that only learns residual parts of CT images and sinograms followed by refinement module;

(d): Prior-Im: network with disentanglement module followed by only image refinement branch;

(e): Prior-Sino: network with disentanglement module followed by only sinogram refinement branch; and

(f): Our full model with both disentanglement and refinement modules.

\begin{figure}[!htb]
\begin{center}
\centering\includegraphics[width=0.5\textwidth]{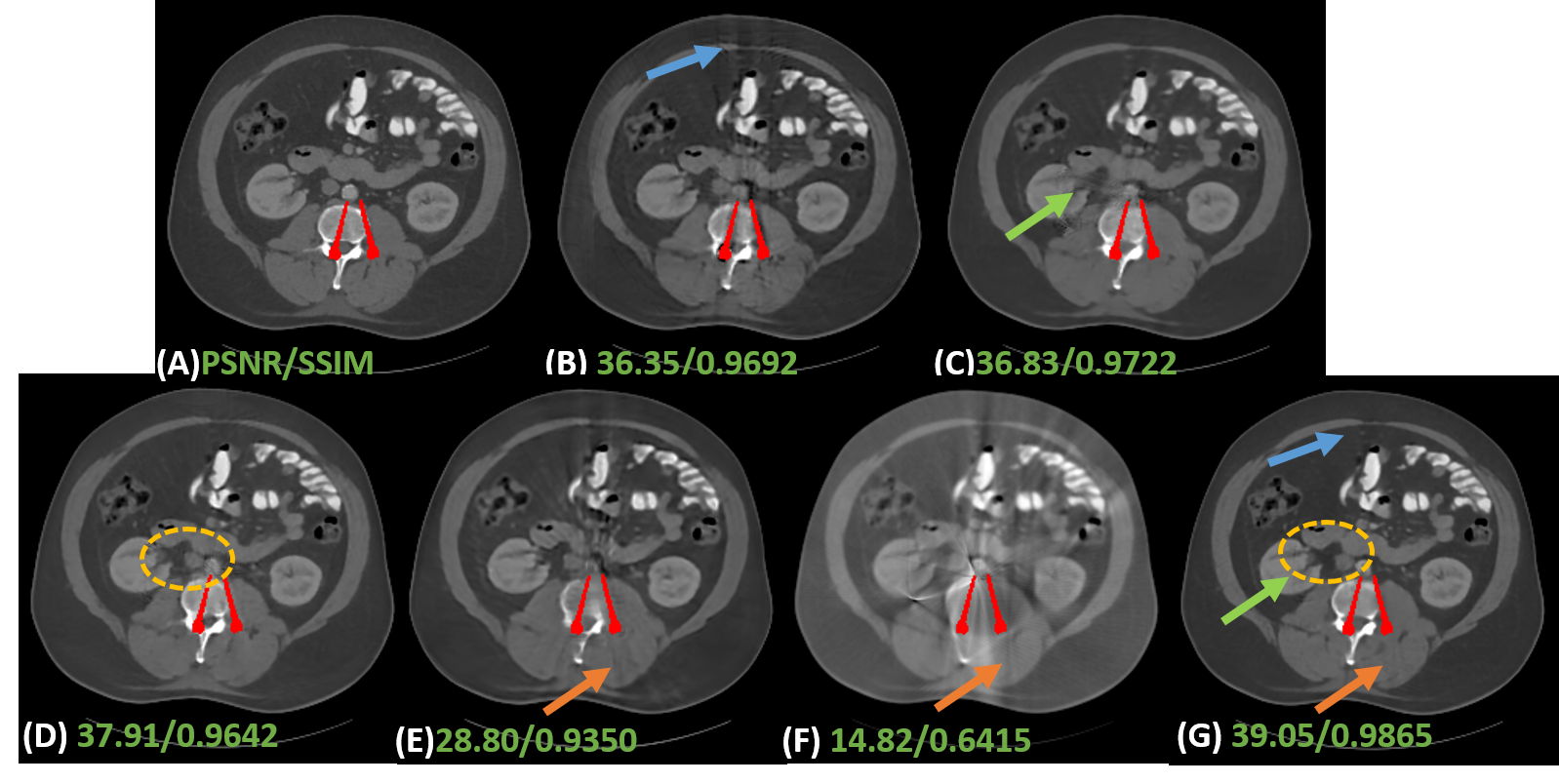}
\end{center}
   \caption{Visual comparison with different MAR methods on a clinical CT image. (A)-(G) respectively represent ground truth and corrected results using Refine-Net, Clean-IDOL, Art-IDOL, Prior-Im, Prior-Sino and IDOL-Net.}
   \label{fig7}
\end{figure}

\textbf{1) Effectiveness of prior estimation.}
Most dual-domain MAR methods usually use original or LI-interpolated sinogram and corresponding reconstructed CT images as inputs, or adopted LI’s corrected results as inputs. However, the first case results in incomplete artifact reduction and maintains some obvious artifacts in final corrected images. In addition, since LI is just a rough estimation, which gives rise to details lost around the metal. A possible solution is to generate both good quality prior sinogram and image. To investigate the effectiveness of the prior image and sinogram generated by our proposed disentanglement module, a network without this module, termed Refine-Net, is trained and we take metal-affected CT image and sinogram as inputs. Table 2 shows the quantitative results. It is clear to see that aided by prior image and sinogram, our model achieves much better score in terms of both metrics. As shown in In Fig. \ref{fig7} (B), there are obvious artifacts indicated by a blue arrow.

\textbf{2) Effectiveness of \(\mathcal{L}_{ma}\).}
There are two ways to obtain high quality prior image and sinogram. One is just predicting the clean data directly from contaminated data. The other is learning residual part and get the clean part by subtracting the learned residual data from original corrupted data. However, these methods may fail to separate the clean data or residuals from original data completely. In other words, there could be some useful information remaining in the restored data or residuals. To tackle this problem, inspired by disentangling representation, we learn both clean part and residual part simultaneously. In addition, we adopt \(\mathcal{L}_{ma}\) to produce a lossless decomposition. To investigate the effectiveness of \(\mathcal{L}_{ma}\), we trained two networks, Clean-IDOL and Art-IDOL, without \(\mathcal{L}_{ma}\). It can be seen in Table 2 that both Clean-IDOL and Art-IDOL achieve lower SSIM and PSNR values than IDOL-Net. In Fig. \ref{fig7}, Clean-IDOL introduced secondary artifacts indicated by green arrow. Art-IDOL shows slight anatomic structure distortion marked by yellow dotted box. It is noticeable that the result of IDOL-Net is most consistent with the referenced image in all the methods.

\textbf{3) Effectiveness of parallel branches.}
To show the effectiveness of our proposed interactive dual-domain parallel branches, we trained two networks, dubbed Prior-Im and Prior-Sino, which are only with image domain branch or sinogram domain branch, respectively. It can be seen in Table 2 that Prior-Sino gains lower SSIM and PSNR scores than other methods. The possible reason lies in that Prior-Sino restored projection data within metal trace region using the original contaminated data and the distortion introduced by inaccurate result will expand to the whole reconstructed image, which heavily lower both quantitative scores calculated in image domain. Prior-Im significantly improves the qualitative and quantitative results compared to Prior-Sino, but as shown in Table 2 and Fig. \ref{fig7}, our IDOL-Net have superior performance to Prior-Im, which provides impressive support for our claim that the proposed interactive dual-domain parallel branches are efficient.

\textbf{4) Discussion of refinement blocks.}
To sense the impact of the number of network blocks in refinement module, different block numbers (from 1 to 5) are tested. TABLE 3 lists the quantitative results and it can be observed that the model with four blocks attains the best quantitative score. 

\begin{table}[]
\centering
\label{table2}
\scalebox{0.9}{
\begin{tabular}{@{}cccccccc@{}}
\toprule
\hline
Methods & 1 Block & 2 Block  & 3 Block &4 Block   &  5 Block \\ 
\midrule
\hline
PSNR & 37.70 & 39.40  &39.34  & \textbf{40.20} &39.37\\
SSIM &0. 9794 & 0.9827 & 0.9831 &\textbf{0.9858}   &0.9808  \\
\bottomrule
\hline
\end{tabular}
}
\caption{The proposed method with different numbers of refinement blocks.}
\end{table}

\section{Conclusion}
In this work, we present a novel interactive dual-domain parallel network for CT MAR, which contains two main modules. The disentanglement module aims to generate high quality prior sinogram and image as the initial inputs for the following refinement module, which is composed of two interactive parallel branches to exchange the useful information extracted from the intermediate results of both sinogram and image domains. The whole framework is trained in a end-to-end manner so that the dual-domain architecture can not only benefit from each other but also alleviate information loss. Experiment results on simulated and clinical data demonstrate that our proposed method can effectively suppress most artifacts and preserve tissues details. In the future work, we will investigate how will the segmentation result influence the final  MAR result. Integrating the segmentation task into our MAR framework is also an interesting topic.

{\small
\bibliographystyle{ieee_fullname}
\bibliography{idol}
}

\end{document}